# Observation of Single Top Quark Production

Ann Heinson

Department of Physics and Astronomy, University of California, Riverside,
Riverside, California 92521; ann.heinson@ucr.edu

Thomas R. Junk

Fermi National Accelerator Laboratory, Batavia, Illinois 60510; trj@fnal.gov

**Abstract**

The field of experimental particle physics has become more sophisticated over time, as fewer, larger experimental collaborations search for small signals in samples with large components of background. The search for and the observation of electroweak single top quark production by the CDF and DØ collaborations at Fermilab's Tevatron collider are an example of an elaborate effort to measure the rate of a very rare process in the presence of large backgrounds and to learn about the properties of the top quark's weak interaction. We present here the techniques used to make this groundbreaking measurement and the interpretation of the results in the context of the Standard Model.

# Contents

## 1. INTRODUCTION

The top quark is the most massive known elementary particle and is produced in very high-energy collisions of other particles. With a lifetime of only $5.2 \times 10^{-25}$ s (1), it decays nearly instantaneously into lighter particles, a *W* boson and a bottom quark (2). By all accounts, it appears to fit well in the Standard Model scheme as a very heavy up-type quark. Its charge is $+(2/3)e$ (3, 4), it couples via the strong interaction in the same way as other quarks (1), and it decays via the weak interaction. Despite this level of knowledge, puzzles remain within the neat classification of particles in the Standard Model. For example, three generations of fermions are known to exist, and there are no more than three flavors of light neutrinos (5). There is no known reason, however, why there should only be three generations. A fourth may exist (6), and possibly more (7), but the new particles' masses and couplings must be such that they have negligible effects on quantities measured precisely at colliders so far. One such measurement is presented in this article – the observation of electroweak production of single top quarks, which constrains a coupling constant of the weak interaction that in turn constrains the character of a fourth generation of fermions.

It is also unknown why the quarks have the masses they have. The Higgs mechanism (8–12) provides an unverified explanation of why quark and gauge boson masses are not zero, but the masses of the quarks are arbitrary parameters that must be measured experimentally before predictions can be made. Since the top quark is the most massive particle, it couples to the Higgs field the most strongly. It could be through observing interactions of the top quark that we can learn something new about what lies beyond our current understanding.

According to Standard Model predictions, top quarks are produced at the Tevatron proton-antiproton collider most often in pairs via the strong force, as shown in **Figure 1**. Indeed, it is in this mode that the top quark was discovered at the Tevatron in 1995 (13, 14). They can also be produced singly via the electroweak interaction in processes shown in **Figure 2**. The Feynman diagrams in **Figure 2 (a)** show the dominant *t*-channel process (15–21), the one in **Figure 2 (b)** shows the lower rate *s*-channel process (21–23), and **Figure 2 (c)** illustrates *tW* production (21, 24), which is not observable at the Tevatron, but is important at the Large Hadron Collider because of the higher energies of its beams.

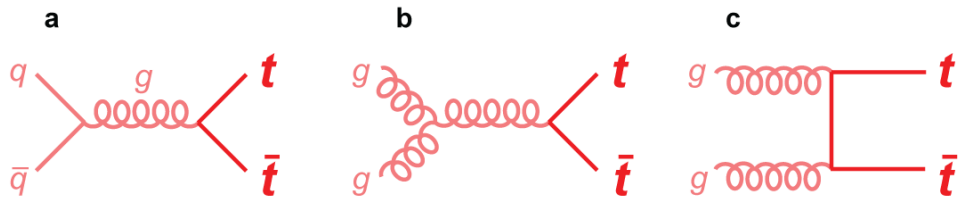

**Figure 1**

Leading-order Feynman diagrams for top quark pair production. The quark-initiated process shown in (*a*) occurs about 85% of the time at the Tevatron and the gluon-initiated processes in (*b*) and (*c*) form the remaining 15%.

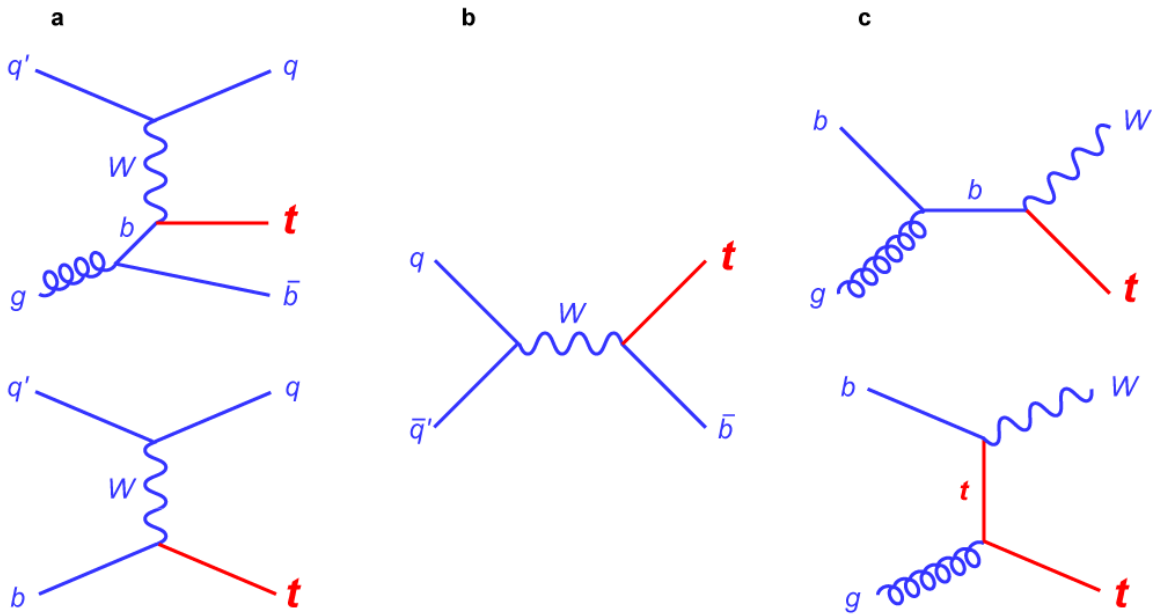

**Figure 2**

Leading-order Feynman diagrams for single top quark production at the Tevatron. (*a*) shows the *t*-channel process, (*b*) shows the *s*-channel process, and (*c*) shows the *tW* process.

Since the strong interaction cannot change the flavor of any particle, a top quark produced by it must be accompanied by a top antiquark. The weak interaction can change one kind of particle into another, and thus it may produce one top quark at a time. The strong force is the stronger one, but the requirement of enough energy to produce two top quarks suppresses the production cross section. Weak production needs only enough energy to produce one top quark (or one top antiquark), but since the strength of the weak interaction is less, the overall cross section is lower. The predicted rates for $t\bar{t}$ production, and for *t*-channel, *s*-channel, and *tW* single top quark production at the Tevatron collision energy of 1.96 TeV are $7.27^{+0.76}_{-0.85}$ pb (25), $2.26 \pm 0.12$ pb , $1.04 \pm 0.04$ pb , and $0.28 \pm 0.06$ pb, respectively (26), for a top quark mass of 172.5 GeV (27).

The electroweak production rate is proportional, for both *s*-channel and *t*-channel production, to the square of the magnitude of the element $V_{tb}$ of the Cabibbo-Kobayashi-Maskawa (CKM) matrix (28, 29), which describes the mixing between quarks to get from the mass eigenstates to the weak-interaction ones. The *Wtb* vertex, whose coupling strength is parametrized by $V_{tb}$, also plays a predominant role in the decay of the top quark. The magnitude of $V_{tb}$ is expected to be very nearly unity in the Standard Model with only three quark generations and left-handed couplings of the *W* boson to fermions. A top quark can also decay into a *W* boson and a strange quark, or a *W* boson and a down quark, but these modes are predicted to occur in fewer than 0.2% of the decays. Even if $V_{tb}$ is quite a bit smaller than 1, the *Wb* decay of the top quark is still expected to be overwhelmingly dominant.

The experimental signatures of *s*-channel and *t*-channel single top quark production are somewhat different from each other. In *s*-channel production, two jets originating from *b* quarks are found in the detector, while in *t*-channel production, the second *b* quark has low transverse momentum and sometimes travels along the beam direction, outside of the acceptance of the detectors. For the *t*-channel, the jet recoiling from the top quark is initiated by a light-flavored quark. CDF and DØ can therefore use these and other characteristics to separate these signal components from each other and from the copious background processes that mimic single top quark production. The *t*-channel and *s*-channel mechanisms are also differently sensitive to physics beyond the Standard Model (30). For example, if $V_{ts}$ is larger than expected, it would enhance the *t*-channel production rate but not the *s*-channel rate. Alternatively, if a *W'* boson exists that plays a similar role as the Standard Model *W* boson in *s*-channel production but is much heavier, it would enhance the *s*-channel production rate much more than that of the *t*-channel process.

The searches for single top quark events at the Tevatron are daunting, much more so than the top quark pair searches sixteen years ago. There are two reasons for this: the production cross section expected for the signal is lower, and the signature of a single top quark event is less distinct from the backgrounds than is $t\bar{t}$ production, as there is only one very massive top quark present in the event and not two. This latter feature requires the collaborations to seek most of the signal in the lepton (electron or muon) plus missing transverse energy (from the neutrino from the *W* decay, which escapes the experimental apparatus undetected) plus two jets final state, while $t\bar{t}$ analyses typically select events with four reconstructed jets. Background events with two jets in them are produced much more often, by approximately two orders of magnitude, than four-jet events. Add to that the fact that mostly only one identified *b* jet is present instead of two, which means that one of the best handles for separating the signal from the background is weaker in single top searches.

Nonetheless, opportunities exist in the single top quark search which can be used to our advantage. The top quark's mass is known, it has just one dominant decay mode, and the properties of *B* hadrons and *W* bosons are well measured. The kinematics of the single top quark production and the system it recoils against are known at next-to-leading order (31–47) and beyond (26, 48), and many detailed Monte Carlo programs exist to simulate it with particle kinematics that match next-to-leading-order calculations: SINGLETOP (49) based on COMPHEP (50), MADEVENT (51) based on MADGRAPH (52), ONETOP (20), ALPGEN (53), TOPREX (54), ACERMC (55), ZTOP (40), MCFM (41), MC@NLO (46), and POWHEG (56). In addition, top quarks produced singly are expected to be nearly 100% polarized along a particular axis (20), which can also be used to form discriminating variables to separate signal events from background ones. Finally, the jets in single top events are almost all from quarks, whereas many of the jets in background events are from gluons, and we use the widths of the jets to separate signal from background, since gluon jets are wider than quark ones (57). The single top quark search benefits from a detailed knowledge in advance of what the signal is expected to look like in the detector, whereas more general searches for new physics must test many hypotheses, seeking particles with fewer distinguishing features. For example, the mass of the Higgs boson is unknown, and it is a scalar particle, giving its decay products a uniform angular distribution.

In order to seek single top quark events, CDF and DØ must therefore be able to reconstruct leptonically decaying *W* bosons, for which good geometric coverage for electrons and muons and good resolution on missing transverse energy are necessary. Jets must be reconstructed out to large values of pseudorapidity ($=\eta=-\log_e[\tan(\theta/2)]$) where $\theta$ is the angle between the jet axis and the beamline), in order to capture the ones that recoil when single top quarks are produced in the *t*-channel.

The observation results (58-60) presented in this paper are extensions of earlier searches (61–67) and evidence papers (68–70) using smaller datasets by the CDF and DØ collaborations.

## 2. COLLECTING AND RECONSTRUCTING DATA

The Tevatron collider is a storage ring in which protons and antiprotons collide at a center-of-mass energy of 1.96 TeV. The CDF (71) and DØ (72, 73) detectors are located in separate interaction regions and collect data from the collisions. A subset of the sensitive elements of the detectors is read out on every beam crossing, which occur at 396 ns intervals. Data are stored in analog and digital pipelines awaiting a trigger decision. Dedicated hardware-based trigger algorithms process the initial data and select events for further consideration at a rate of a few kHz. Once an event is selected at this level, digitization and readout take place, and a second level of trigger decision is applied, reducing the trigger rate to 800 Hz. Both collaborations then use large farms of computers to run fast versions of the offline reconstruction algorithms, the results of which are used to reduce the final trigger rate to 200 Hz and the selected events are permanently recorded. Several different kinds of trigger requirements are applied in parallel in these chains. The trigger algorithms used in CDF's main analysis require the presence of a highly energetic electron or muon. CDF's non-leptonic search (74) requires large missing transverse energy and jets. DØ uses a logical OR of almost all triggers to select its events.

The recorded events are processed offline, producing a calibrated reconstruction of the charged particles' paths in the tracking detectors and the energy deposits in the calorimeters. Jets are reconstructed from calorimeter deposits using a cone algorithm with an angular size $\Delta R = 0.4$ for CDF or 0.5 for DØ. Tracks in the tracking detectors are matched with energy deposits in the electromagnetic calorimeters to form electron candidates and with tracks in the muon chambers to form muon candidates. Dedicated $b$-tagging algorithms (71, 75) identify displaced vertices formed by tracks in the cones of reconstructed jets (illustrated in **Figure 3 (a)**), taking advantage of the large mass (5 GeV) and long lifetime (1.5 ps) of $B$ hadrons. The tagging efficiency is about 50% for $b$ jets from top quark decay. These tagging algorithms are not perfect – charm hadrons sometimes form vertices that pass the $b$-tagging requirements, and tracking misreconstructions and nuclear scattering also allow light-flavored jets (from up, down, and strange quarks, or gluons) to pass the $b$-tagging criteria. The charm-jet tag-rate is about 10% and the light-flavored jet tag-rate is about 1%. For double-tagged events, DØ uses a looser cut on its neural network algorithm to gain signal acceptance.

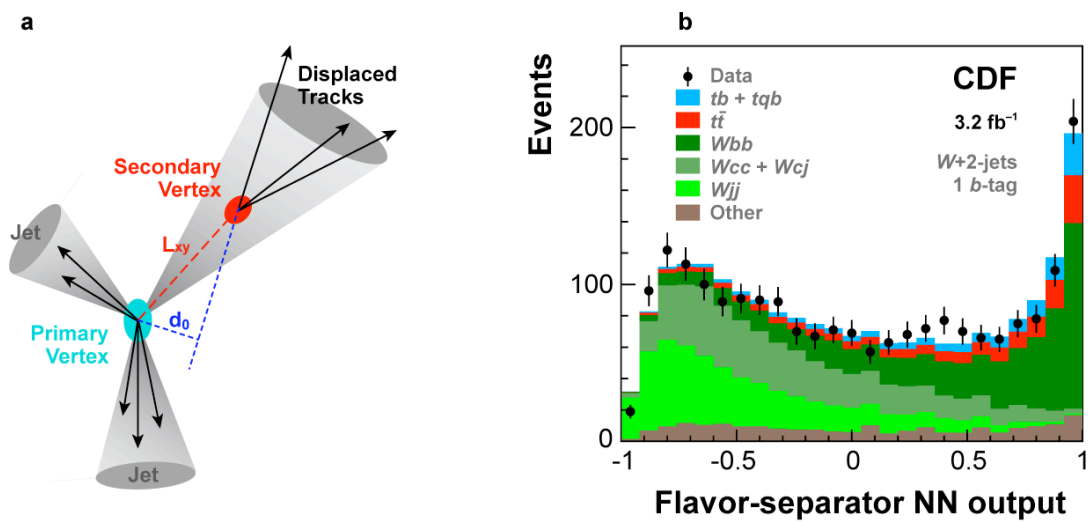

**Figure 3**

($a$) Cartoon of a three-jet event with one identified as a $b$ jet because of the presence of a displaced secondary vertex. $L_{xy}$ is the decay length in the plane transverse to the beamline and $d_0$ is the impact parameter for one of the tracks. ($b$) Distribution of the neural network output for CDF's flavor separator. Events near +1 are most likely to be $b$ jets and events near –1 are most likely to be $c$ jets or light-flavor jets.

CDF applies a neural network flavor separator (60) to *b*-tagged jets to help identify charm and light-flavored jets. The reconstructed secondary vertices' kinematic parameters supply the inputs to the network. Some examples of variables are the invariant mass of the displaced vertex, the highest and second-highest impact parameter significance of any track in the displaced vertex, the decay length of the displaced vertex, and the transverse momentum of the tracks with respect to the jet axis. The neural network also takes advantage of identified electrons or muons within the jet cone. A distribution of the flavor separator neural network output is shown in **Figure 3 (b)**.

## 3. SELECTION OF SIGNAL EVENTS

The first step in analyzing reconstructed events is to select a subsample of them that retains as much of the expected signal as possible while rejecting as many of the background events as possible. These requirements conflict, as some classes of background share many properties with the expected signal. The compromise is to perform a selection that removes easily rejected background events while retaining the signal, and then to use multivariate techniques for further separation of the signal from the background.

The selection of the events is motivated by the properties of the expected signal and backgrounds. Therefore, before we describe the selection, we first explain our model of the signal and describe the properties of the major backgrounds. The CDF experiment uses the MADEVENT Monte Carlo event generator to simulate both the *s*-channel and *t*-channel signals. MADEVENT simulates the final-state partons; PYTHIA (76) is used to simulate the parton showering of the jets and add the underlying event. A detailed detector simulation based on GEANT (77) follows. The DØ experiment uses the SINGLETOP generator, also with PYTHIA parton showering and a GEANT detector simulation. Both MADEVENT and SINGLETOP are tree-level matrix-element generators, and the *t*-channel modeling in particular requires generation of 2→2 events using the 5-flavor diagram and the 2→3 gluon-initiated diagram of **Figure 2 (a)**. Events from both samples are used, based on the momentum of the recoiling *b* antiquark (51). The switchover point is chosen so that the kinematic distributions from the combined leading-order Monte Carlo samples match as closely as possible those from next-to-leading-order calculations (40). The kinematics of the *s*-channel prediction are similar at leading order and next-to-leading order (40) and no adjustment is required. The parton-level kinematics of *t*-channel single top production are shown in **Figure 4**.

One of the major backgrounds in this analysis is *W*+jets, particularly those events in which one or more jets contains a bottom or charm hadron – we call these events *W* + heavy flavor. The *W* + charm events are further split into whether they have one (*Wc*) or two charm quarks ($Wc\bar{c}$) in them – events in which only one charm quark is present are possible since they originate from strange quarks of the sea inside the proton or antiproton and the interaction with the *W* boson can change their flavor to charm. Another significant background is $t\bar{t}$ events (**Figure 1**), and smaller backgrounds include *Z*+jets, diboson events (*WW*, *WZ*, and *ZZ*), and events with no electroweak bosons, called multijet events. Feynman diagrams for the main non-top background processes are illustrated in **Figure 5**.

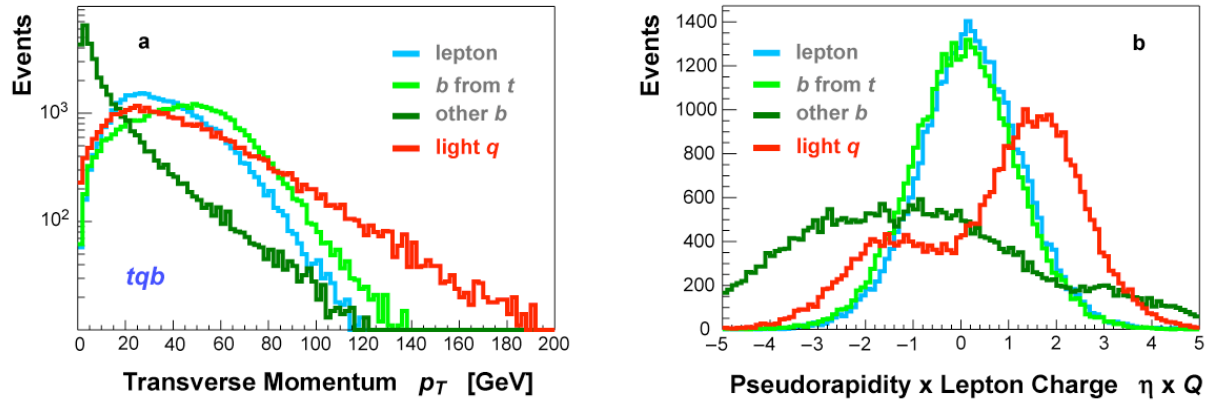

**Figure 4**
Parton-level distributions for the *t*-channel single top quark process after matching to reproduce next-to-leading-order kinematics. (*a*) shows the transverse momentum and (*b*) shows the pseudorapidity times electron or muon charge.

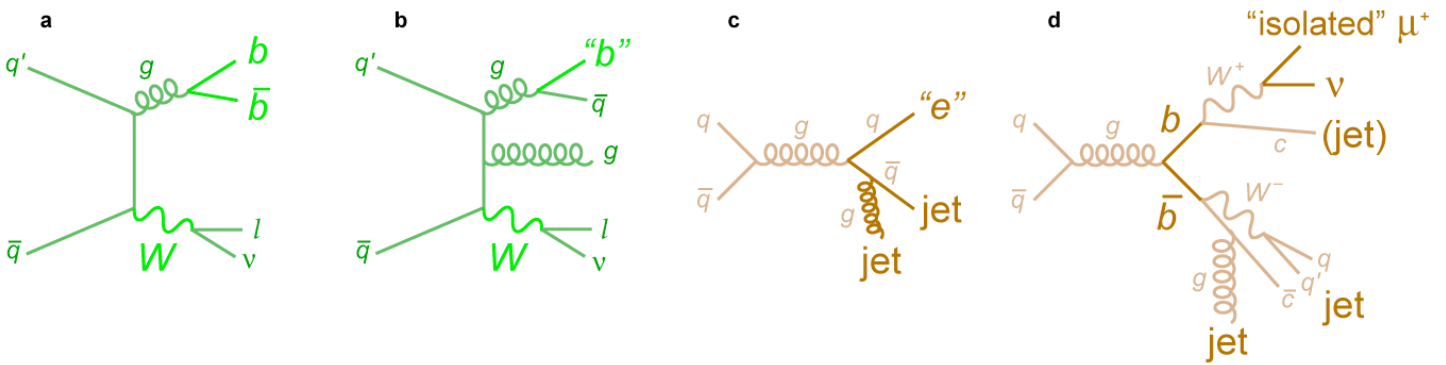

**Figure 5**
Leading-order Feynman diagrams for the main non-top background processes. (*a*) shows the $Wb\bar{b}$ process, (*b*) shows the $Wjjj$ process with a fake *b* tag, (*c*) shows a multijet event where a jet is misidentified as an electron, and (*d*) shows a $b\bar{b}$ multijet event where a muon from a *b* decay travels wide of its jet or the jet does not pass the selection cuts.

We select events with a high-momentum electron or muon plus large missing transverse energy in order to obtain a sample with leptonic *W* boson decays, rejecting nearly all of the multijet events. The remaining multijet events have both misidentified leptons and mismeasured missing transverse energy. In *s*-channel signal events, the top quark decays to *Wb* and recoils against a $\bar{b}$. We therefore expect two *b* jets in these events. In *t*-channel events, the top quark recoils against a light-flavored quark. Jets may fail to be reconstructed if they have too little energy, or travel too close to the beamline. Most of the expected signal events have two or three quarks from the hard scatter, plus possibly one or more radiated gluons. We therefore require events to have two, three, or four jets, and split our data into categories based on the jet counts, since the backgrounds vary from one jet category to another. CDF and DØ apply the selection requirements shown in **Table 1** to reduce the dataset from over a billion events to a few thousand. The cuts for 2-jet events are shown; those for 3-jets and 4-jets are similar but with slightly higher missing transverse energy and total transverse energy requirements to reject the higher multijet background. After all selections, CDF keeps 1.8% of *s*-channel events and 1.2% of *t*-channel events. DØ has looser cuts and keeps 3.7% of *s*-channel events and 2.5% of *t*-channel events. The *t*-channel acceptance is lower than for the *s*-channel for both experiments because the *t*-channel's $\bar{b}$ jet produced with the top quark usually has very low transverse momentum and travels near to the beamline, and so is often not reconstructed.

The vast majority of the jets in *W*+jets events do not contain heavy-flavored hadrons, while signal events always have at least one *b* quark from the top quark decay. We therefore use the *b*-tagging information to help purify the signal sample, since before *b* tagging, the signal-to-background ratio is about 1:260. We separate events into categories that have one *b*-tagged jet and two *b*-tagged jets, as *t*-channel signal events are expected to have predominantly one *b*-tagged jet and *s*-channel signal events have two *b* jets, one or both of which may be *b* tagged. The $t\bar{t}$ background forms a larger fraction of the double-tagged sample, and the *W* + light flavor contribution is substantially reduced in the double-tagged sample as both jets must be mistakenly tagged, which happens less than 0.02% of the time. We use the events in which none of the jets is *b* tagged as a control sample to validate the modeling and to estimate the *W*+jets event rates.

**Table 1** Selection cuts used to identify events that look like signal and reject backgrounds.

|  | CDF's selection | | DØ's selection |
| --- | --- | --- | --- |
|  | **Lepton + 2 jets** | **$\not{E}_T$ + 2 jets** | **Lepton + 2 jets** |
| Electron | $p_T > 20$ GeV | Vetoed | $p_T > 15$ GeV |
|  | $|\eta| < 1.6$ |  | $|\eta| < 1.1$ |
| Muon | $p_T > 20$ GeV | Vetoed | $p_T > 15$ GeV |
|  | $|\eta| < 1.6$ |  | $|\eta| < 2.0$ |
| Neutrino | $\not{E}_T > 25$ GeV | $\not{E}_T > 50$ GeV | $\not{E}_T > 20$ GeV |
| Jet 1 | $p_T > 20$ GeV | $p_T > 35$ GeV | $p_T > 25$ GeV |
|  | $|\eta| < 2.8$ | $|\eta| < 0.9$ | $|\eta| < 3.4$ |
| Jet 2 | $p_T > 20$ GeV | $p_T > 25$ GeV | $p_T > 15$ GeV |
|  | $|\eta| < 2.8$ | $|\eta| < 2.8$ | $|\eta| < 3.4$ |
| Total $E_T$ |  |  | $H_T(\text{jets}, e, \not{E}_T) > 120$ GeV |
|  |  |  | $H_T(\text{jets}, \mu, \not{E}_T) > 110$ GeV |

# 4. BACKGROUND MODEL

In order to observe the production of single top quarks, we must be able to exclude with high confidence the explanation that the data consist only of background events. We therefore have to understand in detail the rates and the kinematic properties of the background processes that contribute events that survive our selection requirements. Our knowledge of the rates and shapes of distributions of observables must come from data not passing the selections, from theoretical predictions, and from Monte Carlo simulations, and this knowledge is input to the statistical procedures described later.

Both collaborations simulate the $t\bar{t}$, $W$+jets, $Z$+jets, and diboson backgrounds with ALPGEN and/or PYTHIA and the full detector simulation. The multijet backgrounds are modeled using data events that pass all selection cuts except that they fail the lepton identification criteria. The $t\bar{t}$, $Z$+jets, and diboson backgrounds are normalized to next-to-leading-order theory cross sections. The $W$+jets background and multijet backgrounds are normalized to untagged (CDF) or pretagged (DØ) data where there is little expected signal, after subtraction of the other background components. **Figure 6** shows the relative fractions of each background source separated by the number of jets and the number of $b$ tags.

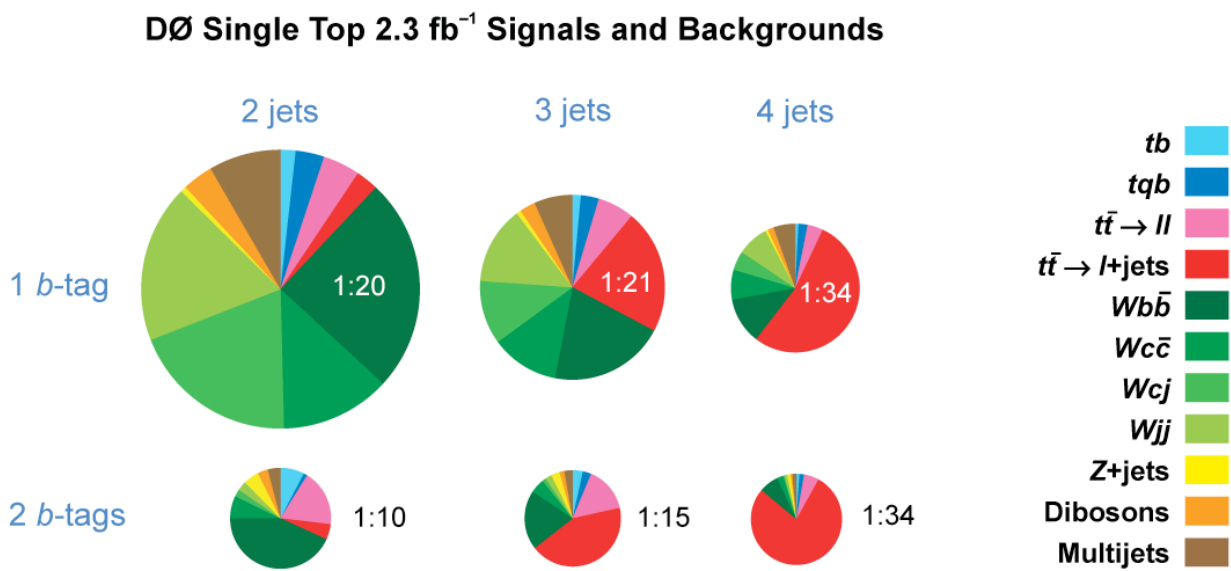

**Figure 6**

Pie charts illustrating the relative fractions of each background source in six decay channels. The area of each pie is proportional to the numbers of events passing the selection. For the most important 2-jet/1-$b$-tag channel, $W$+jets (green shades) dominates the background and the signal:background ratio is 1:20. When exactly two $b$-tagged jets are required, the main backgrounds are $Wb\bar{b}$ (dark green) and top pairs decaying to dileptons (magenta) and the signal-to-background ratio is 1:10. As the number of jets increases, the background is increasingly dominated by top quark pairs decaying to lepton+jets (red).

One important part of the background model is the fraction of heavy flavor jets in W+jets events. The ALPGEN generator simulates this at leading-log precision, giving 0.4% of $Wb\bar{b}$ and 1.1% of $Wc\bar{c}$, but next-to-leading-order theory predicts that these fractions should be 47% higher (78). Similar corrections are needed for the $Wc$ fraction and for the heavy flavor fractions in Z+jets events. Both CDF and DØ measure correction factors from data to account for this limitation of the Monte Carlo event model. The uncertainty on these correction factors is one of the largest components of the total systematic uncertainty in the cross section measurements. CDF applies its heavy flavor separator to the W+1jet data, fitting the fractions of W + light jets, $Wc$, $Wc\bar{c}$, and $Wb\bar{b}$, and then extrapolates the results to the analysis channels with more jets. DØ applies $K'$ factors from the next-to-leading-order calculations, and scales the result by an empirical factor of 0.95 determined from the heavy-flavor fractions in untagged data events.

To identify $b$ jets in Monte Carlo events, several methods are used. CDF applies a tagging algorithm directly to jets originating from $b$ or $c$ quarks. For light quark jets and gluon jets, and for all jets in DØ's event samples, tag-rate functions are applied. These represent the probability that a jet of each type passes the algorithms, as functions of a number of variables, for example in CDF the transverse energy of the jet, the number of tracks it has, the angle from the beam axis to the jet, the number of primary vertices in the event, and the sum of the transverse energies of all the jets. Because any of the jets in an event may pass the tagging algorithm, the collaborations consider each combination of jets passing or failing the algorithm for each Monte Carlo event parametrized by the tag-rate functions.

## 5. EVENT YIELDS

The predicted yields of the expected signal and the backgrounds are listed in **Table 2**. These few thousand events are used to make the final measurements. To compare columns in the table, note: (*i*) CDF uses 175 GeV and DØ uses 170 GeV for the top quark mass, with different theory cross section calculations, which both affect the single top and $t\bar{t}$ expected yields; (*ii*) DØ's analysis includes events with four jets and CDF's does not, so $t\bar{t}$ pairs form a larger fraction of DØ's background than CDF's; and (*iii*) CDF's missing-$E_T$+jets analysis includes W + light jets in the multijet prediction, not in the W+jets prediction.

Distributions of the W boson transverse mass, defined as $M_T(W) = M_T(l, \nu) = \sqrt{2 p_T(l) \not{E}_T (1 - \cos(\phi(l) - \phi(\not{E}_T)))}$ for these events before and after $b$ tagging are shown in **Figure 7**. The symbol $\not{E}_T$ stands for missing transverse energy. All analysis channels have been combined in the plots for DØ. CDF's distribution is just for the main channel, events with two jets and one $b$ tag. Good agreement is seen between the background model plus expected signal and data in each case.

**Table 2** Numbers of events after all selections have been applied.

|  | CDF's yields | | DØ's yields |
|---|---|---|---|
|  | Lepton + jets 3.2 fb$^{-1}$ | $E_T$ + jets 2.1 fb$^{-1}$ | Lepton + jets 2.3 fb$^{-1}$ |
| $tb+tqb$ signal | 191 ± 28 | 64 ± 10 | 223 ± 30 |
| $W$+jets | 2,204 ± 542 | 304 ± 116 | 2,647 ± 241 |
| $Z$+jets, dibosons | 171 ± 15 | 171 ± 54 | 340 ± 61 |
| $t\bar{t}$ pairs | 686 ± 99 | 185 ± 30 | 1,142 ± 168 |
| Multijets | 125 ± 50 | 697 ± 28 | 300 ± 52 |
| Total prediction | 3,377 ± 505 | 1,403 ± 205 | 4,652 ± 352 |
| Data | 3,315 | 1,411 | 4,519 |

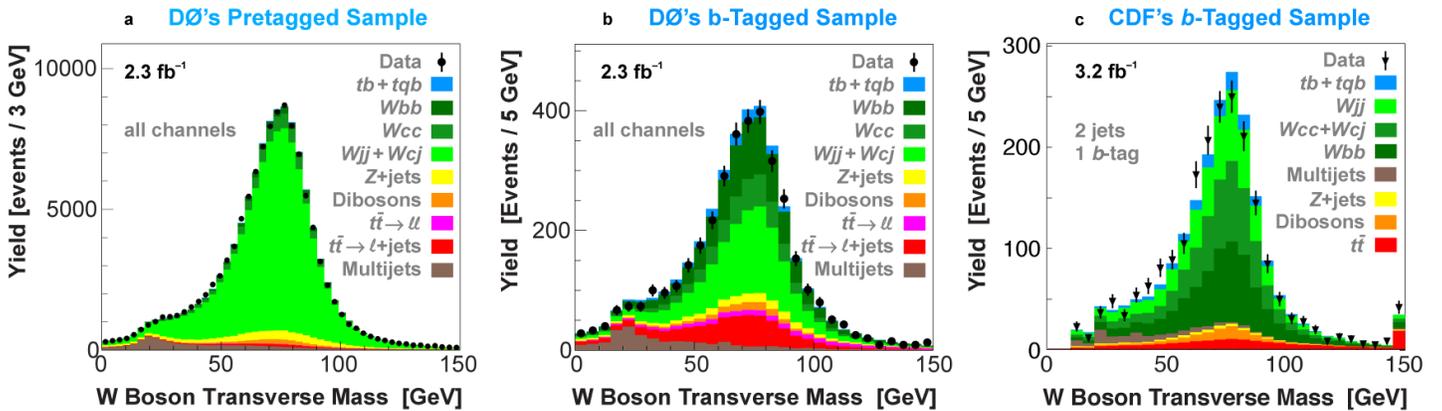

**Figure 7**

Distributions of the transverse mass of the reconstructed $W$ boson for (*a*) DØ with all analysis channels combined before $b$ tagging, where the background is dominated by $W$ + light jets (light green), and (*b*) DØ with all analysis channels combined after $b$ tagging, where the background is evenly divided between $W$+jets (with or without a real $b$ jet) and top quark pairs. (*c*) shows CDF's data with two jets and one $b$ tag, where the background is mostly $W$+jets, split between mistagged light jets and heavy-flavor jets.

# 6. SYSTEMATIC UNCERTAINTIES

The uncertainties on the background model and the signal acceptance are important ingredients for the calculation of the significance of the signal and its cross section. Great care is taken to understand all of the components contributing to these uncertainties. Here, we describe these sources of systematic uncertainty, in order of their impact on the cross section precision.

The largest source of systematic uncertainty is from the modeling of $b$ tagging for Monte Carlo events. The tag-rate functions are measured using several different sets of data and differences found are used to determine the uncertainty as a function of jet transverse momentum and pseudorapidity. The uncertainties therefore apply both to the overall normalization and to the shapes of the distributions.

The second largest source of uncertainty is from the calibration of the jet energy scale. The absolute energies of the jets are determined using many datasets as a function of jet transverse energy and pseudorapidity, and the uncertainties affect both normalization and shape of the background distributions and signal acceptance.

The third source of uncertainty is from the correction to the Monte Carlo model for the fraction of heavy flavor jets produced with $W$ bosons. CDF's method of measuring the correction factor yields an uncertainty of 30% and DØ's method has a 14% uncertainty.

The next set of contributions to the total signal cross section uncertainty are: the integrated luminosity measurement, 6%; smearing functions to correct the modeling of the jet energy resolution, 4%; the modeling of initial-state and final-state radiation, 1–13%; $b$-jet fragmentation modeling, 2%; the $t\bar{t}$ pair production rate using the next-to-next-to-leading-order theory calculation, including the uncertainty on the known and used values of the top quark mass, 13%; and the uncertainty on Monte Carlo correction functions for the electron and muon identification efficiencies, 3%.

There are 14 additional sources of systematic uncertainty identified, measured, and included in the final results. They affect the normalization and are in descending order: a correction factor on the ratio of $Wb\bar{b}$ to $Wc\bar{c}$; primary vertex selection; Monte Carlo statistics; jet fragmentation modeling (we compare results using PYTHIA for creating jets with those from HERWIG (79)); the $W$ boson branching fractions; the $Z$+jets heavy flavor correction; jet reconstruction and identification; the instantaneous luminosity correction; the parton distributions functions (CDF uses CTEQ6.5L (80), DØ uses CTEQ6M (81)); the $Z$+jets theory cross sections; normalization of the $W$+jets and multijets rates using data; and the diboson theory cross sections. All have a small effect on the outcome. Finally, there are sources of uncertainty that affect only the shapes of the distributions, not the normalizations. These are correction functions for the jet pseudorapidity and dijet angle distributions in $W$+jets events in the ALPGEN event generator, the factorization and renormalization scale choices in ALPGEN, the flavor composition of the non-$W$ events, and the trigger turn-on curve correction functions for Monte Carlo events. All systematic uncertainties are propagated through the analysis techniques to the final results, as shown in **Figure 8**.

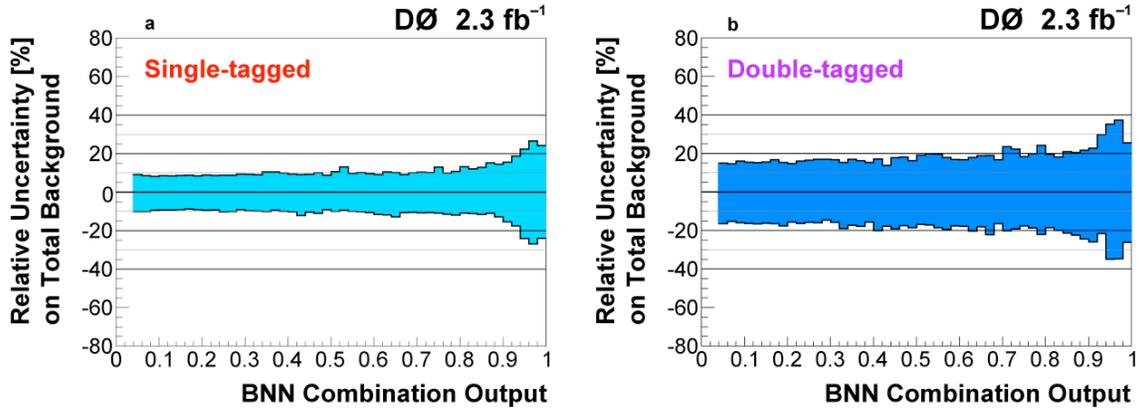

**Figure 8**

Relative uncertainty on the background as a function of the final discriminant output for events with (*a*) one *b*-tagged jet, and (*b*) two *b*-tagged jets, for DØ's analysis.

## 7. ADVANCED ANALYSIS TECHNIQUES

If, after all event selections have been applied, the expected signal-to-background ratio were 10:1 or better, and if sufficient data events are selected, it would be straightforward to declare observation of a signal and measure the cross section by counting events. However, since our average signal-to-background ratio across all analysis channels is about 1:20, which is smaller than the a priori fractional uncertainty on the background rate, this is not possible. Multivariate discriminants need to be applied to separate signal from background before likelihood calculations can be used on the resulting output distributions. This need was anticipated from the start of the analyses, since loose selection cuts were chosen to maximize the signal acceptance, and the analysts knew many background events would also be selected. If tighter selections had been made to maximize instead the signal-to-background ratio, then there would not be enough signal acceptance to make a significant signal observation until many times more data had been collected.

The expected single top signal differs in many ways from the background processes, both in the kinematics of the reconstructed jets, leptons and missing transverse energy, and in the *b*-tagging properties of the jets. Many reconstructed observables for each event capture the features of the signal and background that are different; a careful selection of such features allows the separation to be optimized. The events populate a very high-dimensional space of useful variables. Advanced analysis techniques, described in a review in this volume (82), are used to reduce the dimensionality to a single output variable per selection channel, typically demarcated by the number of reconstructed jets (two, three, or four), the number of *b* tags (one or two), and for CDF whether the event was collected by the lepton trigger or the missing-$E_\mathrm{T}$ plus jets trigger, or for DØ the lepton flavor (electron or muon) and run period (before the silicon tracker (83) upgrade or after it (84)).

Two kinds of analysis techniques are used, which differ fundamentally in their approach. One approach is to identify variables with different distributions for the signal and background, and combine them with likelihood functions (60), neural networks (DØ has used MLPFIT (85) in the past and CDF uses NEUROBAYES (86)), or decision trees (87). These techniques rely on the background model accurately reproducing the data in every variable used. The decision trees and DØ's neural networks undergo a further sophisticated step in their use in which a large number of iterations of the tree or network are averaged using boosting (88) or a Bayesian Markov chain Monte Carlo technique (89) implemented in the FBM software package (90) to improve the separation and stability of the result. DØ uses the RULEFIT package (91) to rank and then select the variables with the best separation power for the Bayesian neural networks. Boosted decision trees can use an almost unlimited number of variables, as they ignore those that are not useful. The second approach is the matrix element method (92), in which the probabilities for all possible interpretations of every event are calculated.

The variables used as inputs to the first class of discriminators fall into six different categories. The first set is of object kinematics, that is, the transverse momenta and angular properties of each individual object (jet, lepton, or missing transverse energy representing the neutrino) in the event. The second set is of event kinematics, where momentum and energy properties of two or more objects are combined. These variables provide information about the particles (like a $W$ or $Z$ boson or a gluon) that produced the final state objects. An example variable in this category is shown in **Figures 9 (a)** and **(b)**. The third set, used by DØ, contains jet reconstruction properties such as the widths (57) and masses of the jets and the momentum of a muon from a $b$ jet relative to the jet axis. The fourth set of variables is a special case of event kinematics, where a top quark is reconstructed using each jet in the event in turn to determine its properties. The fifth set of variables consists of angular correlations (93–95), that is, the angles between each combination of objects in the event. An example of this type of variable is shown in **Figures 9 (c)** and **(d)**. The sixth set, used by CDF, is the jet flavor separator applied to each jet in the event (see **Figure 3**). Each variable is considered after a comparison of background model with data in all analysis channels is found to be within uncertainties, and then is chosen if it shows a different enough distribution between at least one of the signals ($s$-channel or $t$-channel) and at least one of the backgrounds ($t\bar{t} \to$ dileptons , $t\bar{t} \to$ lepton+jets , $Wb\bar{b}$ , $Wc\bar{c}$ , $Wcj$, $Wjj$, $Z$+jets, dibosons, multijets). **Table 3** shows a selection of the variables with the best separation power between signals and $W$+jets and between signals and $t\bar{t}$ pairs.

Both CDF and DØ analyze their data samples using multiple advanced discrimination techniques. Analysis team members participate in this process both by processing the data and evaluating the signal and background expectations and uncertainties, and also by contributing an advanced analysis technique. In order to preserve the work of all contributors while simultaneously further optimizing the sensitivity of the total analysis, the discriminant outputs are combined to produce super-discriminants. This step improves the sensitivity of the analyses

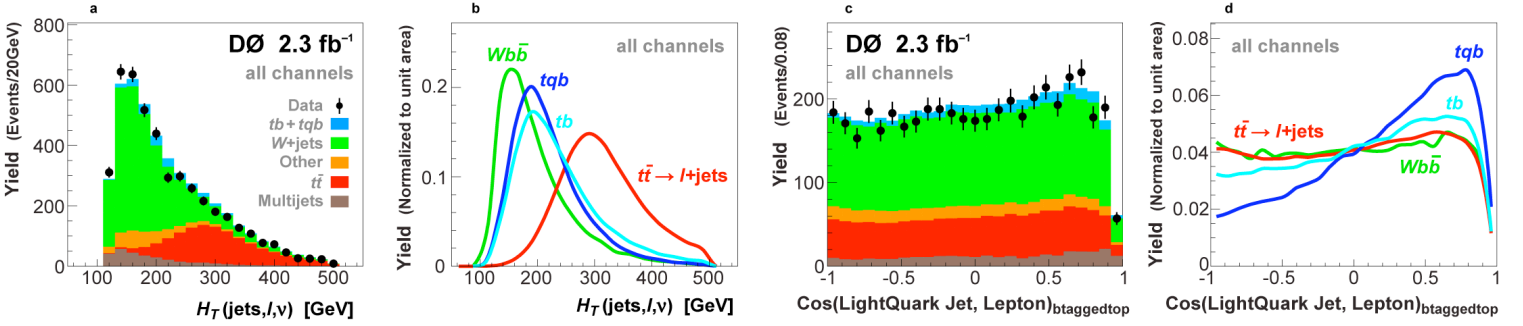

**Figure 9**

Distributions of total transverse energy of all the jets, the lepton, and the missing transverse energy, for (*a*) data with background model after *b* tagging, and (*b*) for the two signals and two main backgrounds with unit area normalization. Plots (*c*) and (*d*) show similar distributions for the angular correlations variable cosine of the angle between the light-quark (untagged) jet and the lepton in the rest frame of the reconstructed top quark using the *b*-tagged jet.

**Table 3**   30 of DØ's 97 variables that have the best signal:background separation.

| Variable type | Separate single top from: | |
|---|---|---|
| | **W+jets** | **$t\bar{t}$ pairs** |
| Object kinematics | $\not{E}_T$ | $p_T$(notbest2) |
| | $p_T$(jet2) | $p_T$(jet4) |
| | $p_T^{\text{rel}}$(jet1, tag-$\mu$) | $p_T$(light2) |
| | $E$(light1) | |
| Event kinematics | $M$(jet1, jet2) | $M$(alljets – tag1) |
| | $M_T$(lepton, $\not{E}_T$) | Centrality(alljets) |
| | $H_T$(lepton, $\not{E}_T$, jet1, jet2) | $M$(alljets – best1) |
| | $H_T$(jet1, jet2) | $H_T$(alljets – tag1) |
| | $H_T$(lepton, $\not{E}_T$) | $H_T$(lepton, $\not{E}_T$, alljets) |
| | | $M$(alljets) |
| Jet reconstruction | Width$_\phi$(jet2) | Width$_\eta$(jet4) |
| | Width$_\eta$(jet2) | Width$_\phi$(jet4) |
| | | Width$_\phi$(jet2) |
| Top quark reconstruction | $M_{\text{top}}$(W($\nu$-solution1), tag1) | |
| | $\Delta M_{\text{top}}^{\min}$ | |
| | $M_{\text{top}}$(W(($\nu$-solution2), tag1) | |
| Angular correlations | cos(light1, lepton)$_{\text{btaggedtop}}$ | cos(lepton$_{\text{btaggedtop}}$, btaggedtop$_{\text{CM}}$) |
| | $\Delta\phi$(lepton, $\not{E}_T$) | $Q$(lepton) × $\eta$(light1) |
| | $Q$(lepton) × $\eta$(light1) | $\Delta R$(jet1, jet2) |

since the outputs of the separate methods are correlated by 57%–85%. If they were 100% correlated, then no improvement could be achieved. For this combination, CDF uses a neural network with genetic evolution known as NEAT (96) and DØ uses a set of Bayesian neural networks (90). The figure of merit on which CDF's super-discriminant is optimized is related to the expected signal significance. DØ's networks are optimized in a more traditional manner using an error function, which measures the signal-background similarity. We use the entire output distributions of the super-discriminants to extract the results, even though the signal-to-background ratio is high enough to discern the signal in only the uppermost bins. The high-background bins constrain the systematic uncertainties in situ. CDF's $\not{E}_T$+jets channels analyze an independent dataset from the lepton+jets data and thus are combined as a statistically independent set of channels rather than as part of the super-discriminant. The output distributions for the super-discriminants for all channels combined are shown in **Figure 10.**

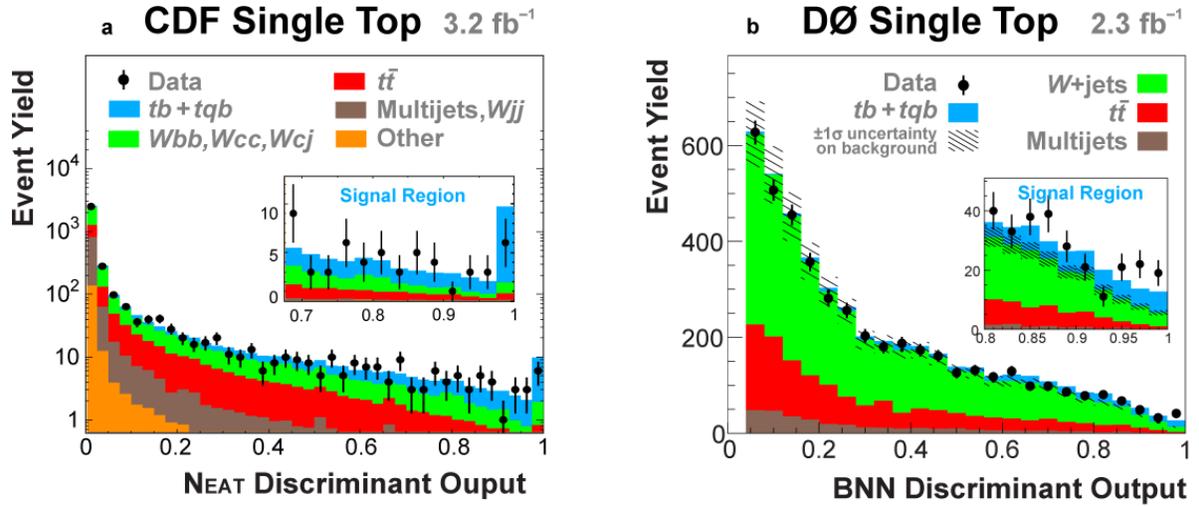

**Figure 10**

Output distributions for (*a*) CDF's NEAT super-discriminant and (*b*) DØ's Bayesian neural networks super-discriminant, for all channels combined.

## 8. SIGNAL SIGNIFICANCE

The two goals of the single top analysis are to be able to establish the presence of a signal at high significance (if a signal is truly present), and to measure the production cross section $\sigma_{s+t} = \sigma(p\bar{p} \to tb + X, tqb + X)$ with as high an expected precision as possible. The significance of the data observed in excess of the background prediction is expressed with a *p* value: the probability that an outcome at least as signal-like as the data can be produced by the background model, where outcomes are ranked using a real-valued test statistic. CDF's choice of test statistic is the likelihood ratio (97) comparing the null hypothesis (the Standard Model with no single top produced), with the test hypothesis (the Standard Model plus single top produced at the Standard Model rate), and is given by

$$-2\log_e Q = -2\log_e \frac{\text{Likelihood}(\text{data} \mid \text{signal+background, nuisances})}{\text{Likelihood}(\text{data} \mid \text{background, nuisances})}.$$

The systematic uncertainties are parametrized with nuisance parameters. Both the numerator and the denominator are separately maximized over the possible values of the nuisance parameters. DØ's choice of test statistic is the measured signal cross section. If the *p* value is less than $2.87 \times 10^{-7}$, then an observation at the 5-standard-deviation level can be claimed.

To compute the *p* values, the CDF and DØ collaborations create ensembles of pseudodatasets that each represent the background expected in the final dataset. The number of events in each pseudodataset is on average equal to the number predicted, but fluctuates for each set with Poisson statistics. The systematic uncertainties are also included in the pseudodatasets with Gaussian distributions and all correlations are taken into account. The expected *p* value for CDF is obtained from the fraction of background-only pseudodatasets with $-2\log_e Q$ no higher than the median value of $-2\log_e Q$ from the signal+background pseudodatasets. DØ's expected *p* value is obtained from the fraction of background-only pseudodatasets that produce a signal cross section at least as high as the Standard Model expected value. The *p* value calculations are illustrated in **Figure 11**.

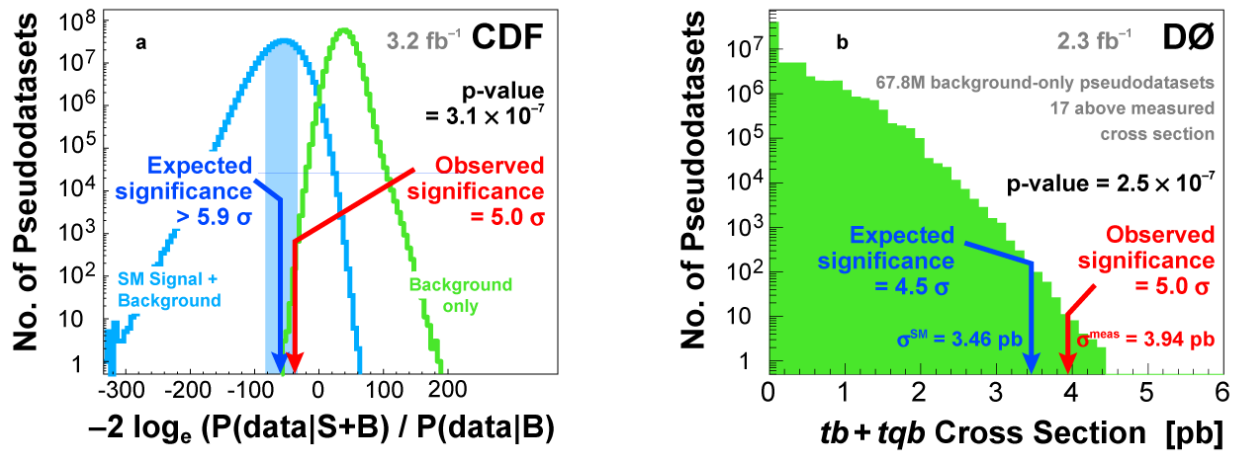

**Figure 11**

Significance measurements for the single top quark signal from (*a*) CDF and (*b*) DØ.

# 9. CROSS CHECKS

We check the background model by defining datasets that contain mostly *W*+jets events or mostly $t\bar{t}$ pairs. This allows both the normalization and shape of these two main background components to be independently verified, as shown in **Figure 12**.

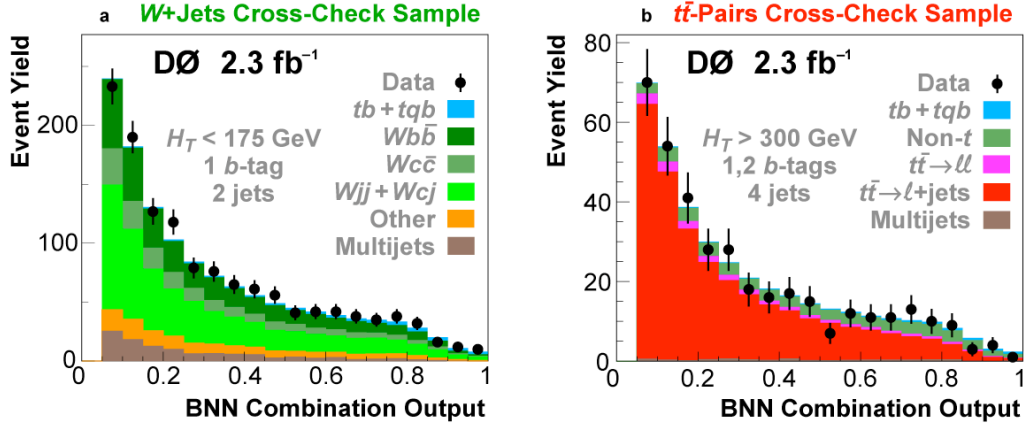

**Figure 12**

Cross-check samples after DØ's Bayesian neural networks super-discriminant for all channels combined, which show good normalization and shape agreement for (*a*) mainly *W*+jets events and (*b*) mainly $t\bar{t}$ pair events.

The linearity of the cross section measurement method is tested by measuring the cross section in ensembles of pseudodatasets that include both background and Monte Carlo signal events. Each ensemble has a different signal cross section. The results of the check are shown in **Figure 13**.

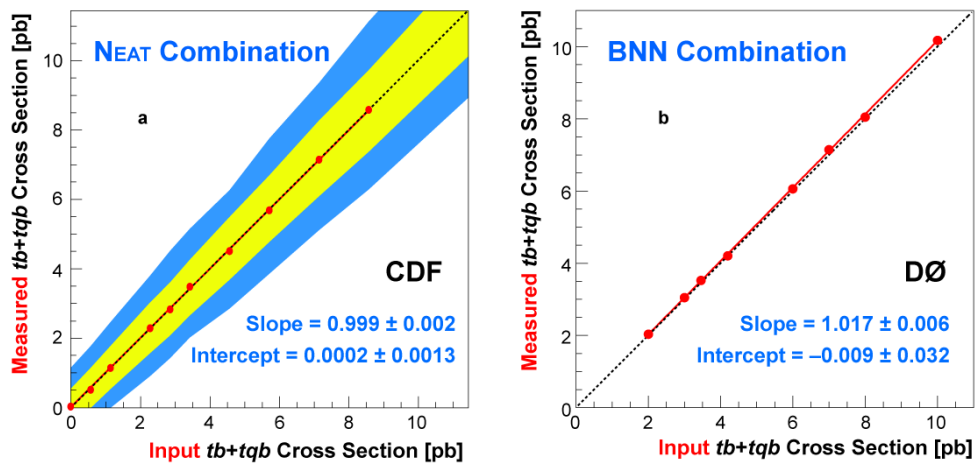

**Figure 13**

Linearity tests of the cross section measurement methods for (*a*) CDF and (*b*) DØ. The CDF plot indicates the distribution of expected outcomes with shaded bands, where the yellow band indicates the 68% confidence interval and the green one indicates the 95% confidence interval. For both plots, the dotted lines show perfect linearity and the red lines show the fits to the points.

# 10. RESULTS

**Total Cross Section Measurement.** The single top quark cross sections are measured using a Bayesian calculation (98) combining the data, the signal acceptance, and the background estimate, with all uncertainties, in every bin of the output distributions of the discriminants in each analysis channel. Thousands of measurements are combined into posterior probability densities from which we get the final cross section results. A uniform non-negative prior for the cross section is assumed. The *s*-channel+*t*-channel cross section measurements and the expected and observed significances for each analysis technique are summarized in **Table 4** for CDF and DØ. The two collaborations have also exchanged the histograms of their data and the corresponding signal and background predictions, with all systematic uncertainties, as well as the programs required to compute the cross sections. A joint posterior was formed, allowing the total cross section to be measured combining the CDF and DØ measurements (99) in the same way that the separate measurements within the collaborations are combined. This measurement is also listed in **Table 4**.

Table 4  Single top quark cross sections and significances for each analysis.

| Analysis | | Single top cross section (pb) | Significance (standard deviations) | |
|---|---|---|---|---|
| | | | Expected | Measured |
| **CDF** | Boosted decision trees | $2.1^{+0.7}_{-0.6}$ | 5.2 | 3.5 |
| | Neural networks | $1.8^{+0.6}_{-0.6}$ | 5.2 | 3.5 |
| | Matrix elements | $2.5^{+0.7}_{-0.6}$ | 4.9 | 4.3 |
| | Likelihoods | $1.6^{+0.8}_{-0.7}$ | 4.0 | 2.4 |
| | Likelihoods, *s*-channel | $1.5^{+0.9}_{-0.8}$ | 1.1 | 2.0 |
| | Combination, lepton+jets | $2.1^{+0.6}_{-0.5}$ | | |
| | Neural networks, $\not{E}_T$ +jets | $4.9^{+2.6}_{-2.2}$ | 1.4 | 2.1 |
| | **Combination (175 GeV)** | $\mathbf{2.3^{+0.6}_{-0.5}}$ | **> 5.9** | **5.0** |
| | **Combination (170 GeV)** | $\mathbf{2.35^{+0.56}_{-0.50}}$ | | |
| **DØ** | Boosted decision trees | $3.74^{+0.95}_{-0.79}$ | 4.3 | 4.6 |
| | Bayesian neural networks | $4.70^{+1.18}_{-0.93}$ | 4.1 | 5.4 |
| | Matrix elements | $4.30^{+0.99}_{-1.20}$ | 4.1 | 4.9 |
| | **Combination (170 GeV)** | $\mathbf{3.94 \pm 0.88}$ | **4.5** | **5.0** |
| **Tevatron combination (170 GeV)** | | $\mathbf{2.76^{+0.56}_{-0.47}}$ | | |
| Theory ($M_{top}$ = 170 GeV) | | $3.46 \pm 0.18$ | | |

DØ has estimated the components of its total cross section uncertainty of 22% that come from data statistics and from the combined systematic uncertainties: they are 18% from data statistics and 13% from systematics. The cross section measurements are therefore just moving from being statistics-limited towards being limited in their precision by the knowledge and control of the systematic uncertainties.

$|V_{tb}|$ **Measurement.** The total single top quark production cross section is proportional to $|V_{tb}|^2$, neglecting components that are proportional to $|V_{ts}|^2$ and $|V_{td}|^2$ since they are expected to be very much smaller. If the 3×3 Cabibbo-Kobayashi-Maskawa (CKM) matrix is unitary, then $|V_{tb}|$ is nearly unity, but if the CKM matrix is either larger, owing to the presence of a fourth generation of quarks or more, or if it is not unitary for other reasons, then $|V_{tb}|$ can be smaller (30, 100, 101). If $|V_{tb}|$ is suppressed owing to new physics, the dominant top quark decay mode remains $t \rightarrow W^+ b$. We therefore extract a measurement of $|V_{tb}|$ from our single top quark cross section results (18, 21) using the relationship

$$|V_{tb}|^2_{\text{measured}} = |V_{tb}|^2_{\text{theory}} \times \frac{\sigma^{\text{measured}}_{s+t}}{\sigma^{\text{theory}}_{s+t}}.$$

We use the same technique as we used for the measured cross section to form a posterior density in $|V_{tb}|^2_{\text{measured}}$, although in this case we also include the uncertainty on $\sigma^{\text{theory}}_{s+t}$. Two posteriors are formed: one is chosen only to be non-negative, $|V_{tb}|^2 \geq 0$, while the other is further constrained to lie within the physical region $0 \leq |V_{tb}|^2 \leq 1$. The first choice is used to obtain a measurement of $|V_{tb}|$ that can be averaged with other measurements in an unbiased way, while the second one is used to find a physically allowed range. The combined results (99) from CDF and DØ assuming $\sigma^{\text{theory}}_{s+t} = 3.46$ pb and $|V_{tb}|_{\text{theory}} = 0.999$, are

$$|V_{tb}|_{\text{measured}} = 0.88 \pm 0.07,$$

and $\quad 0.77 < |V_{tb}|_{\text{measured}} \leq 1$ at the 95% confidence level.

**Separate *t*-Channel and *s*-Channel Measurements.** CDF and DØ also measure the *t*-channel and *s*-channel cross sections separately, using a two-dimensional extension of the technique used for the combined cross section. A uniform prior is assumed in the $(\sigma_s, \sigma_t)$ plane, allowing only for non-negative cross section measurements. Independent information for these two cross sections comes largely from the separation of the data into single-*b*-tagged and double-*b*-tagged subsets, although the kinematic differences between the two signals contribute even within the analysis channels. The two-dimensional results are shown in **Figure 14**. The numerical values are

CDF (60)    $\sigma_t = 0.8\ ^{+0.4}_{-0.4}$ pb       $\sigma_s = 1.8\ ^{+0.7}_{-0.5}$ pb,

DØ (102)    $\sigma_t = 3.14\ ^{+0.95}_{-0.80}$ pb   $\sigma_s = 1.05 \pm 0.81$ pb.

The theory values for comparison are $\sigma_t = 2.34 \pm 0.12$ pb and $\sigma_s = 1.12 \pm 0.04$ pb (26). CDF's result is not in good agreement with the SM prediction; the disagreement is visible as a deficit of data with respect to the signal plus background prediction in the single-*b*-tagged channels and an excess of data in the double-*b*-tagged channels (60). DØ's *t*-channel measurement is one standard deviation higher than the theory and has a significance of 4.8 standard deviations, almost at the observation level, and the *s*-channel result is very close to the theory prediction.

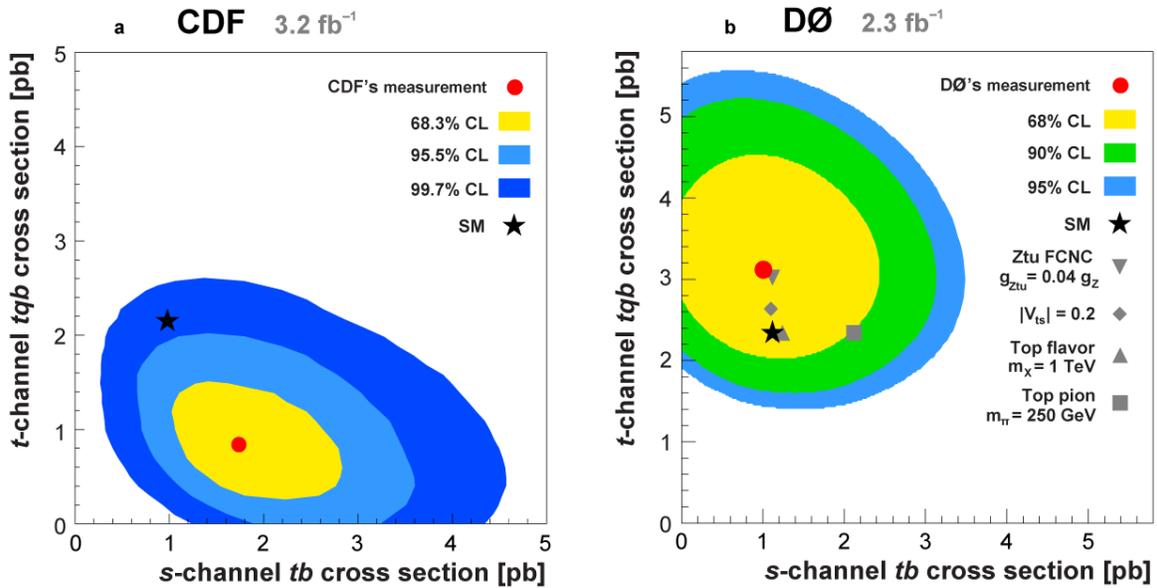

**Figure 14**

Simultaneous measurements of the *s*-channel *tb* and *t*-channel *tqb* cross sections from (*a*) CDF and (*b*) DØ, together with the theory values and some beyond-the-Standard-Model predictions (16). CDF's measurements are for a top quark of mass 175 GeV and DØ's for 170 GeV. The theory cross sections shown are from Kidonakis 2006 (14).

## 11. WHAT'S NEW SINCE THE OBSERVATION?

**Tau Decay.** DØ has published a search for single top in the decay channel $t \to Wb$, $W \to \tau\nu$, $\tau \to$ hadrons using 4.8 fb$^{-1}$ of data (103). The final state therefore has no electron or muon in it, just missing transverse energy and jets. A new more efficient algorithm for hadronic $\tau$ identification in events with other jets has been developed for this analysis. Although a signal is not observed, the new upper limit on single top production of 7.3 pb at 95% confidence level is combined with the observation result to get an improved cross section measurement: $\sigma_{s+t}(p\bar{p} \to tb+X, tqb+X) = 3.84^{+0.89}_{-0.83}$ pb.

**Heavy Flavor Resonances.** Several models of physics beyond the Standard Model predict new particles that behave like $W$ bosons but are heavier (30, 104–106). Following on from earlier searches for $W' \to tb$ (107–110) and $H^{\pm} \to tb$ (111), DØ has a new result (112) setting higher mass limits on a proposed heavy $W'$ boson that decays to $t\bar{b}$ or $\bar{t}b$, ruling out new particles up to about 900 GeV in several models.

**Anomalous Couplings.** Although single top is not very sensitive to anomalies in the strong $gt\bar{t}$ coupling (113), it is the process that is the most sensitive to the electroweak $Wtb$ coupling (21, 114–118). One can also search for flavor-changing neutral-current couplings $tgu$ and $tgc$ (119–125). Previously, DØ published two results looking for anomalous $Wtb$ couplings in single top quark production (126, 127). The second result combines the single top measurement with the $W$ helicity one from top decay in $t\bar{t}$ pairs (128). Nothing anomalous is observed, and limits are set on all combinations of pairs of couplings, left-handed or right-handed, vector or tensor. Flavor-changing neutral current couplings $tgu$ and $tgc$ have also been searched for. DØ's measurements use events with a lepton and missing transverse energy and at least two jets, one $b$ tagged (129), and CDF's measurement uses similar events but with exactly one jet, $b$ tagged (130). DØ has a new measurement that sets tight limits on the anomalous coupling coefficients and also on the branching fractions $B(t \to gq)$ and cross sections $\sigma_{tgq}$, where $q$ is an up or charm quark (131).

**Top Quark Width and Lifetime.** Theorists have proposed that the rate of single top quark production may be used to determine the decay width of the top quark and hence its lifetime (132, 133). The predicted values at next-to-leading-order precision are 1.26 GeV and $5.2 \times 10^{-25}$ s respectively (1). DØ has recently performed a combination of its $t$-channel single top cross section measurement (102) with the top quark branching ratio of decays to $Wb$ over $Wq$, (where $q$ means here any down-type quark: down, strange, or bottom) from $t\bar{t}$ pair decays (134) to determine the top quark width and lifetime. The results are $\Gamma_{\text{top}} = 1.99^{+0.69}_{-0.55}$ GeV and $\tau_{\text{top}} = \left(3.3^{+1.3}_{-0.9}\right) \times 10^{-25}$ s (135). These results are then used to constrain the coupling between the top quark and the $W$ boson and a fourth-generation $b'$ quark for the first time, giving $|V_{tb'}| < 0.63$ at the 95% confidence level.

# 12. PROSPECTS AT THE LARGE HADRON COLLIDER

The Large Hadron Collider (LHC) at the CERN international particle physics laboratory in Switzerland collides protons on protons at a center-of-mass energy of 7 TeV. It started long-term operation in March 2010 and collected ~40 pb$^{-1}$ of data until switching to heavy ion collisions in November 2010. It will restart proton collisions in early 2011, possibly at a slightly higher energy. The design calls for collisions at 14 TeV, which will occur after a long shutdown, restarting in 2013 or 2014. The first studies for single top quark physics at the LHC have been performed using simulations of 14 TeV collisions, see e.g. reference (136) by the ATLAS collaboration or (137) by the CMS collaboration. The theory cross sections at 14 TeV are: *t*-channel, 150 pb for top quarks and 92 pb for top antiquarks; *tW* production, 50 pb for top quarks and 50 pb for top antiquarks; and for *s*-channel, 8 pb for top quarks and 4 pb for top antiquarks (138). The rates for top and antitop are different in the *t*-channel because top is produced from a valence quark in the proton together with a gluon, whereas antitop is produced from a sea quark and a gluon. In the *s*-channel, the asymmetry comes about because a quark and an antiquark are needed from the protons, and a proton has two up valence quarks (charge $+(2/3)e$) and one down valence quark (charge $-(1/3)e$), with sea quarks coming in roughly equal quantities, so $u\bar{d} \rightarrow t\bar{b}$ is more likely than $\bar{u}d \rightarrow \bar{t}b$. The *tW* process starts from a *gb* initial state, with the *b* coming from $g \rightarrow b\bar{b}$, so top and antitop production are equally likely. Once operation of the LHC at 14 TeV starts, there will be very large samples of single top quark events and high precision measurements of the cross sections, $|V_{tb}|$, the top quark width and lifetime, and searches beyond the Standard Model will be performed.

At 7 TeV, the single top cross sections fall to less than half the 14 TeV values. With the very small datasets currently available, searches for single top quark production have not yet been made public. After the nearly catastrophic magnet fault of September 2008, the beam energy was planned to be at 10 TeV, and therefore studies at this energy for 200 pb$^{-1}$ of data have been completed (139, 140). The collaborations select events with a lepton + $\not{E}_T$ + *b*-tagged jets using simple kinematic cuts, and predict 470 *t*-channel events for ATLAS and 102 muon *t*-channel events for CMS, with signal-to-background ratios of 1:8 and 1:2 respectively. The background models use leading-order theory cross sections for normalization, and so are somewhat underestimated. Various tighter selections are then applied to improve the sensitivity, which is expected to give a cross section uncertainty of about 40% (141). Two new studies address single top quark selection and polarization measurements at the LHC, using next-to-leading order modeling (142, 143). They include detailed discussions on many aspects of single top quark physics at the LHC at all three beam energies.

## 13. SUMMARY

Single top quark production in the *t*-channel was proposed in 1986 (15) and in the *s*-channel in 1991 (22). First evidence was found by DØ in 2006 (68, 69) and by CDF in 2008 (70). In 2009, the CDF and DØ collaborations simultaneously announced observation of single top quark production (58–60). Approximately 15 years elapsed between the start of the search and its completion because the signal is small and the process looks similar to the very large backgrounds. Not only did enough data have to be collected, but the tools used to model the backgrounds and to identify the final state objects in the events had to be significantly improved in this period. Pioneering work has been performed in the use of multivariate discriminants to separate signal from background, and the single top quark physics programs are at the leading edge in sophistication in the development and application of experimental techniques in high-energy physics today.

## DISCLOSURE STATEMENT

The authors are supported by U.S. Department of Energy funds to pursue particle physics research at the Fermilab Tevatron collider. AH is a member of the DØ collaboration and TRJ is a member of the CDF collaboration.

## ACKNOWLEDGMENTS

The authors would like to thank the members of the CDF and DØ collaborations, their theoretical colleagues, the Fermilab staff, and the technical staffs of the participating institutions for their vital contributions. Fermilab is operated by Fermi Research Alliance LLC under contract with the U.S. Department of Energy.

# Key Terms/Definitions

| | |
|---|---|
| Tevatron collider: | Proton-antiproton accelerator at Fermilab, 4 miles in circumference, operating at a center-of-mass energy of 1.96 Tera-electron-Volts = $1.96 \times 10^{12}$ eV. |
| Standard Model: | All the known elementary particles and their interactions. |
| Top quark, $t$: | The heaviest known quark, an up-type quark of the third generation, mass = 173.3 Giga-electron-Volts = $173.3 \times 10^9$ eV. |
| Bottom quark, $b$: | The down-type third-generation partner of the top quark, mass = 4.67 GeV. |
| $W$ boson: | Charged carrier of the weak force, mass = 80.399 GeV. |
| $Z$ boson: | Neutral carrier of the weak force, mass = 91.1876 GeV. |
| Gluon, $g$: | Massless carrier of the strong force. |
| Monte Carlo: | Method of simulating particle physics events using probabilities. |
| Discriminator: | Combination of many variables into one, which separates signal from background. |
| $p$ value: | Probability that the background-only prediction fluctuates up to give the signal cross section or higher. |

# Abbreviations/Acronyms

| | |
|---|---|
| Fermilab: | Fermi National Accelerator Laboratory, Batavia, Illinois, USA. |
| CDF: | Collider Detector at Fermilab. |
| DØ: | Collider experiment also at the Fermilab Tevatron; pronounced "D-Zero." |
| CKM: | Cabbibo-Kobayashi-Maskawa quark mixing matrix. |
| $V_{tb}$: | The CKM matrix element that connects the top quark to the bottom quark via the $W$ boson. |
| CERN: | Conseil Européen pour la Recherche Nucléaire (European Laboratory for Particle Physics), Geneva, Switzerland. |
| LHC: | Large Hadron Collider at CERN, 16.8 miles in circumference, colliding protons on protons at a center-of-mass energy of 7 TeV, scheduled to increase to 14 TeV in a few years' time. |
| ATLAS: | A Toroidal LHC Apparatus, an experiment at the LHC. |
| CMS: | Compact Muon Solenoid, an experiment at the LHC. |

## Summary Points

1. The top quark is the heaviest elementary particle of the Standard Model. Discovered in 1995 at the Fermilab Tevatron, it has only been observed until now in top-antitop pairs produced via the strong interaction.
2. Because the top quark is so heavy, it is expected to interact most strongly of all particles with the as yet unobserved Higgs boson, the particle predicted to give mass to all particles via electroweak symmetry breaking.
3. It is important to understand the weak interactions of the top quark, exemplified by weak production of single top quarks, in order to be able to check for new physics beyond the Standard Model.
4. New techniques had to be developed and applied in order to separate the tiny signal from huge backgrounds of similar events. These techniques are now being used in the Higgs boson searches at Fermilab and CERN.
5. The CDF and DØ collaborations simultaneously announced the first observation of single top quark production in March 2009, a milestone in new knowledge about the top quark and in the experimental techniques used to obtain that knowledge.

## Future Issues

1. Single top quark production has been observed using about 3 fb$^{-1}$ of data by the CDF and DØ collaborations at Fermilab. They each now have 9 fb$^{-1}$ of data recorded. A proposed extension to the Tevatron run could double this dataset by 2014. Analyzing more data will reduce the statistical and systematic uncertainties on all the measurements.
2. The *t*-channel and *s*-channel production modes will be observed independently at the Tevatron, allowing tighter constraints to be placed on more processes predicted to occur beyond the Standard Model.
3. The Large Hadron Collider at CERN will collect enough data for single top quark production to be observed in a couple of years' time in the *t*-channel and *tW* production modes. The low-rate *s*-channel mode will need much more data to be separated from background.
4. Other processes not yet studied at the Tevatron, which will become accessible at the Large Hadron Collider include single top quark production via Kaluza-Klein excitations in models with large extra dimensions, technipions, *R*-parity-violating supersymmetric squarks or sleptons, littlest Higgs models with *T*-parity, and *CP*-violation as seen in the bottom and kaon sectors.
5. Single top quarks are also predicted to be produced together with a neutral Standard Model Higgs boson, with a charged Higgs boson from supersymmetric models, with a charged heavy pion, with a heavy neutral scalar particle, or with a charmed quark via flavor-changing neutral-current interactions. These processes can all be searched for at the Large Hadron Collider experiments.

## Annotated References

15. Presents the first description of single top quark production via the *t*-channel.
22. Gives the first description of single top quark production via the *s*-channel.
35. First calculation of the *s*-channel cross section including higher-order corrections.
36. First calculation of the *t*-channel cross section at next-to-leading order.
39. Presents fully differential cross section calculations for *s*-channel and *t*-channel production.
58. The 5-standard-deviation observation of single top quark production on which this review is based, by the DØ collaboration.
59. The 5-standard-deviation observation of single top quark production on which this review is based, by the CDF collaboration.
62. Shows the first application of neural networks to separate signal from background in a high-energy physics analysis, by the DØ collaboration.
68. Presents 3-standard-deviation evidence for single top quark production, by the DØ collaboration.
70. Presents 3-standard deviation evidence for single top quark production, by the CDF collaboration.

## Related Resources

Schwienhost R (2006). Search for Single Top Quark Production at DØ. *Mod. Phys. Lett. A* 21:1339 (2006)

Demina R, Thomson EJ (2008). Top Quark Production and Properties. *Ann. Rev. Nucl. Part. Sci.* 58:125 (2008)

Heinson AP (2010). Observation of Single Top Quark Production at the Tevatron Collider. *Mod. Phys. Lett. A* 25:309 (2010)

## Side Bar 1 – Standard Model Higgs Boson Searches

The single top quark searches are very similar to searches for the Standard Model Higgs boson in the mode $p\bar{p} \to WH + X$, $W \to l\nu$, $H \to b\bar{b}$, since the final-state particles are the same. Single top quark production is therefore a background for the Higgs boson search, and backgrounds in the single top search are also backgrounds to the Higgs search. The Higgs boson is more difficult to find than single top because the predicted cross section is much smaller, the mass is not yet known, and the decay products are isotropically distributed in its rest frame, whereas there is more kinematic structure in the top quark decay. Recent examples of Higgs boson searches use 2.7 fb$^{-1}$ of data by the CDF collaboration (144) and 5.3 fb$^{-1}$ by the DØ collaboration (145), where an upper limit of 4.5 times the predicted rate for a Higgs boson of mass 115 GeV at 95% confidence level has been set. The observation of single top quark production is a proof of principle that a small signal can be extracted from a very large multicomponent background.

## Side Bar 2 – A Prior "Discovery" of the Top Quark

A cautionary tale illustrating the difficulty of this type of analysis is the mid-1980's "discovery" of the top quark produced singly in $W$ boson decays ($p\bar{p} \to W^{\pm} + X$, $W^+ \to t\bar{b}$, $W^- \to \bar{t}b$) in events with a lepton plus missing transverse energy plus jets in the final state, with a measured top quark mass of 40 GeV (146). The $W$+jets background is copious and difficult to understand, and the early effort's background model and uncertainties were not well understood. The new techniques applied by CDF and DØ make extensive use of the data to better characterize this difficult background.